\documentclass[twocolumn, aps, longbibliography]{revtex4-1}
\usepackage{graphicx}
\usepackage{xcolor}
\usepackage[utf8x]{inputenc}

\begin{document}

\title{Thickness-dependence of hydrogen-induced phase transition in MoTe$_{2}$}

\author{Priyanka Manchanda}
\author{Pankaj Kumar}
\author{Pratibha Dev}
\affiliation{Department of Physics and Astronomy, Howard University, Washington, D.C. 20059, USA}

\begin{abstract}

Two-dimensional transition metal dichalcogenides (TMDs) usually exist in two or more structural phases with different physical properties, and can be repeatedly switched between these phases via different stimuli, making them potentially useful for memory devices. An understanding of the physics of interfaces between the TMDs and conventional semiconductors, or other 2D-crystals forming heterogenous or homogeneous assemblies is central to their successful application in technologies.  However, to date, most theoretical works have explored phase-change properties of isolated TMD monolayers in vacuum. Using \textit{ab-initio} calculations, we show how interfacial effects modify the thermodynamics and kinetics of the phase transition by studying hydrogen-induced transitions in monolayers and bilayers of MoTe$_{2}$. The phase-change properties of MoTe$_{2}$ show substantial thickness-dependence, with the timescale for a transition in the hydrogenated bilayer being about $10^7$-times longer than that in a monolayer at room temperature. Our study highlights the importance of taking effects of immediate environment into account when predicting properties of 2D crystals.

\end{abstract}
\maketitle


\section{Introduction}

Phase-change materials, which undergo changes in atomic arrangement under different stimuli, are of interest for various applications, such as memory devices to supplement or replace silicon-based memory, and neuromorphic computing~\cite{Lencer2008, Burr2010, Hayat2017,Pantazi2016,Kuzum2012}. 
Amongst different 2D structures, transition metal dichalcogenides (TMDs) are actively being studied for their phase-change properties. The greatest attention has been paid to the group-VI TMDs, MX$_{2}$ (M = Mo, W; X = S, Se, Te), which adopt more than one crystal structure~\cite{Chhowalla2013,Eda2012,Yang2017}. With the exception of WTe$_{2}$, the lowest energy phase of most of these TMDs is the trigonal prismatic semiconducting $H$-phase. The other two common phases are metallic and include, the octahedrally-coordinated \textit{T}-phase and the $T^\prime$-phase, which is a distorted-\textit{T} structure. 
The TMDs are able to reversibly switch between these phases, making them desirable phase change materials for scaling down memory devices and improving the performance of flexible electronics. 

Within the group-VI-based TMDs, MoTe$_{2}$ is considered to be the most promising candidate for phase change materials due to very small energy difference between the $H$- and  $T^\prime$-phases~\cite{Li2016}. Several experimental and theoretical works have explored different means of attaining phase change, such as strain~\cite{Duerloo2014, Song2016}, chemical modifications~\cite{Cho2015, Zhang2016, Rhodes2017, Young2017, Hwang2017, Nan2019} and gating or electrostatic doping~\cite{Zhang2016, Wang2017, Li2016, Krishnamoorthy2018, Zhang2019}. Most experiments are performed with the TMDs on substrates, which may be conventional 3D crystals such as silicon dioxide, or other 2D-crystals forming heterogenous or homogeneous assemblies. On the other hand, most of the theoretical works utilize freestanding monolayers in vacuum, even though the presence of a substrate (another 2D layer or 3D crystals) can affect the properties of a 2D crystal in unpredictable ways~\cite{Dev_PRB_gr_cu_2014, Josh_Robinson_MoS2_Dope_2015}. Experiments, indeed, indicate that layer thickness and/or substrates influence the phase-change properties and relative phase stability. For example, it was found that MoTe$_{2}$ grown on InAs(111)/Si(111) substrate is stable in the metallic $T^\prime$-phase at room temperature. This reversal in phase stability is attributed to the strain in MoTe$_{2}$ induced by the substrate~\cite{Tsipas2018}.  

In this work, we elucidate the effects of substrates/layer-thickness on phase change properties of MoTe$_{2}$. In order to demonstrate proof of principle and to keep the study focussed, we considered the simple case of MoTe$_{2}$ placed on another MoTe$_{2}$-layer, where the bottom layer serves as the substrate. Hydrogenation was used as a mean of inducing phase change in MoTe$_{2}$ monolayers and bilayers. We chose hydrogen-adsorption due to its reversibility~\cite{Nan2019} and versatility, as it can be used to induce a local phase change through masking, or a global phase transition through the exposure of the entire crystal. In this sense, it is unlike other means such as: (i) electrostatic doping that results in a global phase change, or (ii) the laser-induced creation of defects that results in irreversible phase-change. Although there are several theoretical works that have studied phase change in MoTe$_{2}$ with hydrogen adsorption~\cite{Zhou2015, Qu2016}, the mechanism of phase change, effects of hydrogen concentration and substrates/layer-thickness have not been studied thus far. Using density-functional theory (DFT), we explain the underlying phase-change mechanism and show that it is quite different from the phase-change mechanism(s) involved in the case of alkali-metal adsorption, as well as charge-doping/gating. We further show that the kinetic energy barrier to phase transition is considerably reduced upon hydrogenation of the monolayers of MoTe$_{2}$, while it remains almost unchanged in the case of a bilayer, showing the importance of
taking substrate-effects/layer-thickness into account when predicting properties of 2D crystals.

\section{Calculation Details}

Spin-polarized calculations were performed using projector-augmented wave (PAW) pseudopotentials as implemented in the Vienna Ab-initio Simulation Package (VASP).~\cite{Kresse1996}. The generalized gradient approximation (GGA) of Perdew-Burke-Erzerhof (PBE)~\cite{Perdew1996} was used to approximate the exchange-correlation functional. For the bilayer, van der Waals interactions  (rev-vdW-DF2 method) were included in the calculations~\cite{Hamada2014}. Along with the monoclinic $T^\prime$-phase, MoTe$_{2}$ can also form an orthorhombic structure, which is similar to the monoclinic phase~\cite{Chang2016,Qi2016}. As the energy difference between the two structures is very small ($\sim0.25$\,meV/f.u., with formula units abbreviated as f.u.), we used the orthorhombic structure of the $T^\prime$-phase. We also adopted the rectangular unit cell of the $H$-phase structure, consisting of two formula units. These choices allowed us a direct comparison between the calculations performed for the $T^\prime$-and the $H$-phase structures. All of the reported energies are given in the units of eV/f.u.. In order to model different concentrations of the hydrogenated $H$/$T^\prime$-phase, we constructed a number of rectangular supercells ($6\times4$, $4\times4$, $3\times3$, $4\times2$, $3\times2$, and $2\times2$). The \textit{k}-point mesh density was preserved for supercell calculations, which was equivalent to a $24\times12\times1$-grid for the rectangular unit cell consisting of two-formula units. The hydrogen coverage was calculated by considering that there is only one-sided absorption of hydrogen on Te for both the monolayer and bilayer. For example, when one hydrogen atom is adsorbed on the supercell of a $6\times4$ MoTe$_{2}$ monolayer, it corresponds to 2.08 \% hydrogen concentration. The hydrogenated $H$/$T^\prime$-phase layers were separated by at least 15\,\AA{} of vacuum to eliminate the spurious interactions between the periodic images. The kinetic energy cut-off was set to 600\,eV. The energy convergence criterion was set to 10$^{-6}$\,eV and the atomic relaxations were carried out until the forces were smaller then 10$^{-2}$\,eV/\AA{}. To calculate the transition energy barrier between $H$ and  $T^\prime$, we employed the climbing-image nudged elastic band method as implemented in VASP through VTST tools~\cite{Henkelman2000}.

\begin{figure} [h]
	\centering 
	\includegraphics[width=1.0\linewidth]{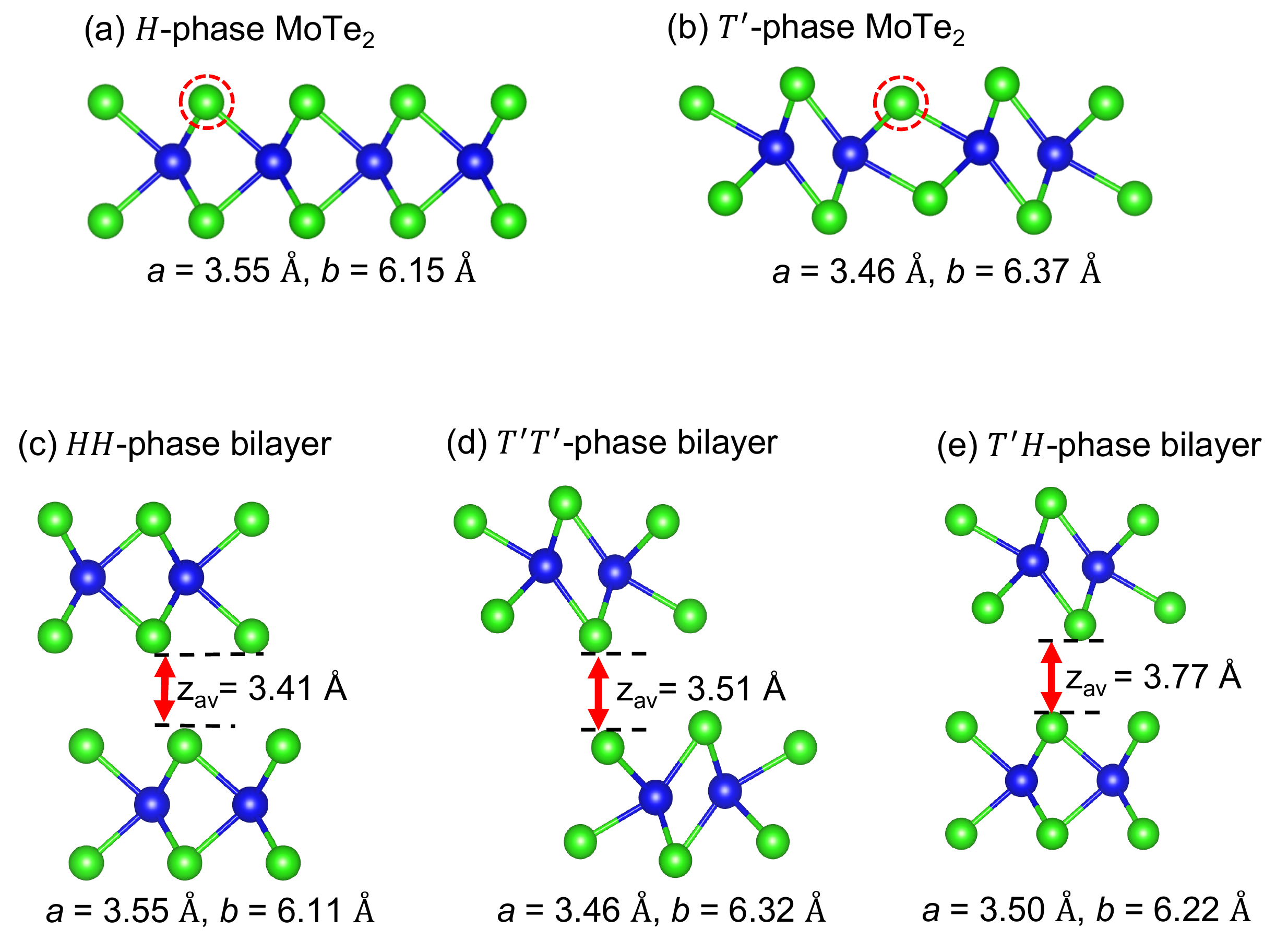}
	\caption{Atomic structures of (a) $H$, (b) $T^\prime$ phase MoTe$_{2}$ monolayers and (c) $HH$, (d) $T^{\prime}T^{\prime}$, (e) hybrid $T^{\prime}H$ phase MoTe$_{2}$ bilayers. The encircled Te atoms in (a, b) are the most favorable hydrogen adsorption sites for $H$ and $T^\prime$ phase, respectively. $a$ and $b$ are the in-plane lattice constants of the unit cells.}
	\label{fig:figure-1}
\end{figure}


\section{Pristine Monolayers and Bilayers}

Before investigating the physico-chemical effects of hydrogenation on MoTe$_{2}$, we calculated the structural and electronic properties of pristine monolayers and bilayers of MoTe$_{2}$. Figures \ref{fig:figure-1}(a) and (b) show the optimized structures for the semiconducting $H$-phase and the semimetallic $T^\prime$-phase of MoTe$_{2}$. The optimized $H$-phase monolayer is a semiconductor with a band gap of 1.06\,eV, and is more stable than the metallic $T^\prime$-phase by 0.039\,eV/f.u., consistent with previous DFT based reports~\cite{Kan2015}.  As a bilayer, MoTe$_{2}$ can exist in three different phases depending upon the different phase combinations of the two vertically stacked layers: (i) both in $H$ phase ($HH$-phase), (ii) both in $T^{\prime}$ phase ($T^{\prime}T^{\prime}$-phase), and (iii) hybrid $T^{\prime}$- on $H$-phase heterostructure ($T^{\prime}H$-phase). The latter structure is included in this study as it is suspected to be an intermediate state in transition from the $HH \longrightarrow T^{\prime}T^{\prime}$-phase structure upon hydrogenation of the top layer within the $HH$-composite. Several high-symmetry configurations are possible with different stacking orders for the bilayers. For each phase, the most energetically favorable stacking sequence is shown in Figs~\ref{fig:figure-1}(c-e).  The $HH$-bilayer favors $AB$-stacking, $T^\prime$$T^\prime$ favors $AA^\prime$-stacking, while $AA$-stacking is favored for the hybrid $T^{\prime}H$-bilayer. 
The average distances between the bilayers are given in Figs~\ref{fig:figure-1}(c-e), with the average distances showing the trend: $z_{av}(HH) <  z_{av}(T^{\prime}T{^\prime}) < z_{av}(T^{\prime}H)$. This trend follows the binding energies of the bilayers, defined as: $\Delta_{B.E.}= E_{bilayer}-\sum_{i}{E^{i}_{monolayer}}$. Here, the summation is over the constituents of the bilayer. The binding energies of the bilayers in the $HH$, $T^{\prime}T^{\prime}$ and $T^{\prime}H$ phases are $-0.256$\,eV/f.u., $-0.235$\,eV/f.u., and $-0.184$\,eV/f.u., respectively.  These are consistent with our results showing that the $HH$-phase bilayer is energetically more favorable than the $T^{\prime}T^{\prime}$-phase bilayer (by 0.071\,eV/f.u.), and the $T^{\prime}H$ bilayer (by 0.097\,eV/f.u.). Hence, we find that the energy differences between different phases for the bilayers are about twice as large as those predicted for the monolayers (0.039\,eV/f.u.). This shows that if we account for effects of substrates/layer-thickness on phase-stability of 2D crystals (here, between the $H$-phase and the $T^{\prime}$-phase of MoTe$_2$), DFT may predict different values than those predicted for freestanding monolayers in vacuum. These differences will have to be considered when designing experiments.


\section{Hydrogenated Monolayers and Bilayers: Phase Transition}

\begin{figure}
	\centering
	\hspace*{-0.5 cm}
	\includegraphics[width=0.35\textwidth]{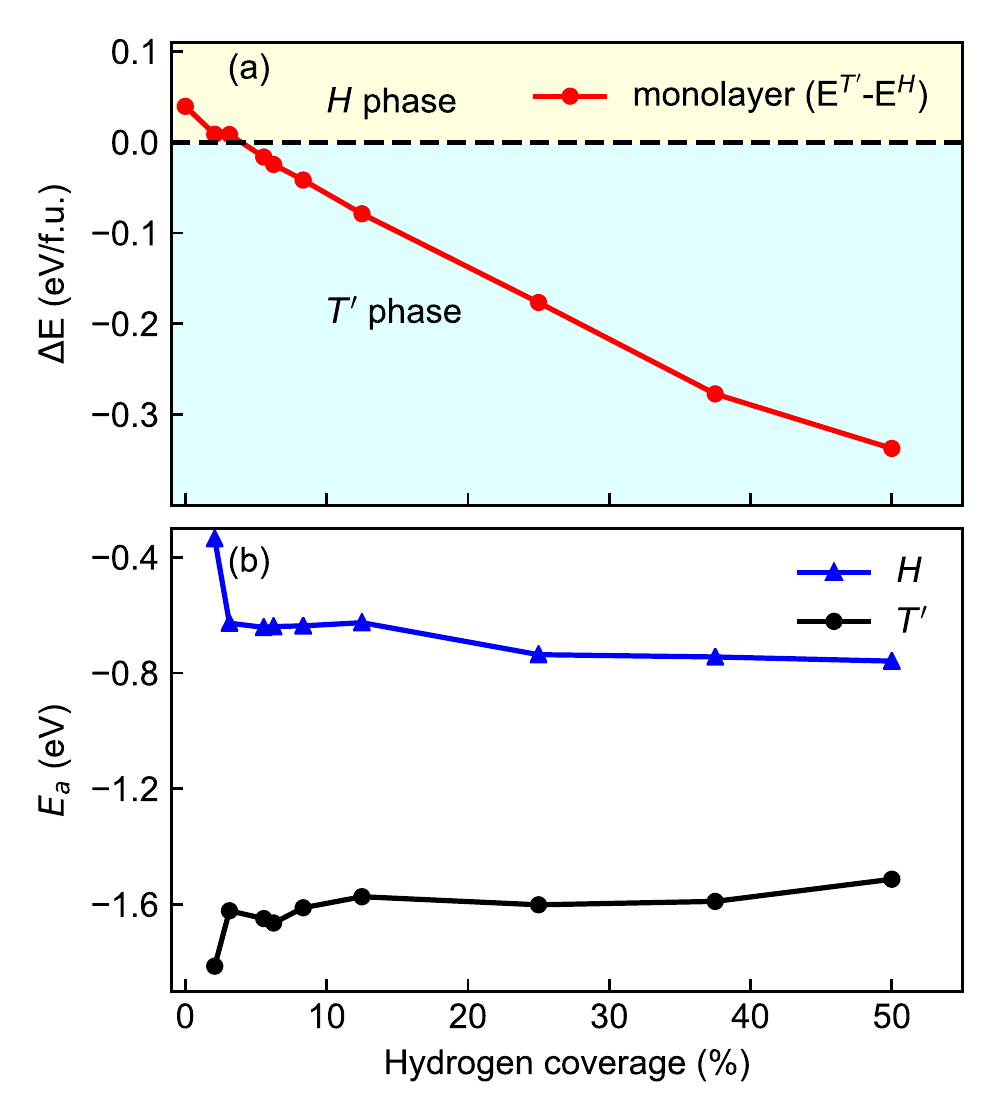}
	\caption{(a) The energy difference between different phases of MoTe$_{2}$ monolayers as a function of hydrogen concentration; (b) adsorption energy as a function of hydrogen concentration for $H$ and $T^\prime$ phases for MoTe$_{2}$ monolayers.}
	\label{fig:phase-transition+adsorb}
\end{figure}

\noindent \textbf{Freestanding  MoTe$_{2}$ Monolayers:} Before investigating how the presence of a second layer may modify the phase-change properties, we studied the hydrogenation-induced phase change in MoTe$_2$ monolayers for a number of adatom coverages. There are several possible sites within the MoTe$_{2}$ layers where a hydrogen atom can be adsorbed. The encircled atoms in Fig. \ref{fig:figure-1}(a,b) indicate the most favorable sites for hydrogen adsorption on the $H$-phase and the $T^\prime$-phase of MoTe$_{2}$.  In order to find the relative stability of the two phases, we calculated the difference in the total energies of the functionalized freestanding MoTe$_{2}$ monolayers in the $T^\prime$ and the $H$-phases, $\Delta E=E^{T^{^\prime}}-E^{H}$, as a function of hydrogen coverage. As illustrated by the change in the sign of $\Delta E$ in Fig.~\ref{fig:phase-transition+adsorb}(a), MoTe$_2$ undergoes a structural phase transition as the hydrogen coverage is increased. In the case of a monolayer, this transition occurs at $\sim4\%$ coverage. In addition, Fig.~\ref{fig:phase-transition+adsorb}(a) shows that the stability of the structure in the $T^\prime$ phase increases monotonically as a function of hydrogen concentration. However, beyond 50\% coverage, the steric repulsion between adsorbed hydrogen atoms starts to dominate, destabilizing the structure and resulting in the spontaneous formation of H$_{2}$ molecules that leave the surface. 
 
To understand the relative stability of the two phases of MoTe$_{2}$ monolayers upon hydrogen adsorption, we considered several inter-related changes in structural and electronic properties upon hydrogen functionalization that may be contributing to the observed trend. The first of these factors is adsorption energy, $E_a$, which gives the strength of the interaction of the adatom with the surface and is defined as:
\begin{equation}\label{equation:eq1}
E_a =1/m[E(\textrm{(MoTe$_{2}$)} _n \textrm{H}_m )-nE(\textrm{MoTe}_2 )-mE(\textrm{H})]	
\end{equation}

Here, $E(\textrm{MoTe}_2) _n \textrm{H}_m$, $E(\textrm{MoTe}_2 )$ and $E(\textrm{H})$ are the total energies of a $H$- or $T^\prime$-phase TMD monolayer containing $n$ formula units of MoTe$_{2}$ with a total number of $m$ adsorbed hydrogen atoms, total energy of a pristine $H$ or $T^\prime$-phase MoTe$_{2}$ monolayer and energy of an isolated hydrogen atom, respectively. The adsorption energies of a hydrogenated $H$ and $T^\prime$ MoTe$_{2}$ monolayer are plotted as a function of hydrogen coverage in Fig. \ref{fig:phase-transition+adsorb}(b). The adsorption of hydrogen is an exothermic process for both the $H$-phase and $T^{\prime}$-phase MoTe$_{2}$ monolayers. The adsorption energies have larger negative values for the $T^{\prime}$-phase at all coverages as compared to the $H$-phase MoTe$_{2}$.  Hydrogen adsorption, therefore, stabilizes the higher-energy phase of MoTe$_{2}$ more than the lower-energy $H$-phase monolayer, shifting the stability of MoTe$_{2}$ towards the $T^\prime$ phase. 

\begin{figure}
	\centering
	\includegraphics[width=0.45\textwidth]{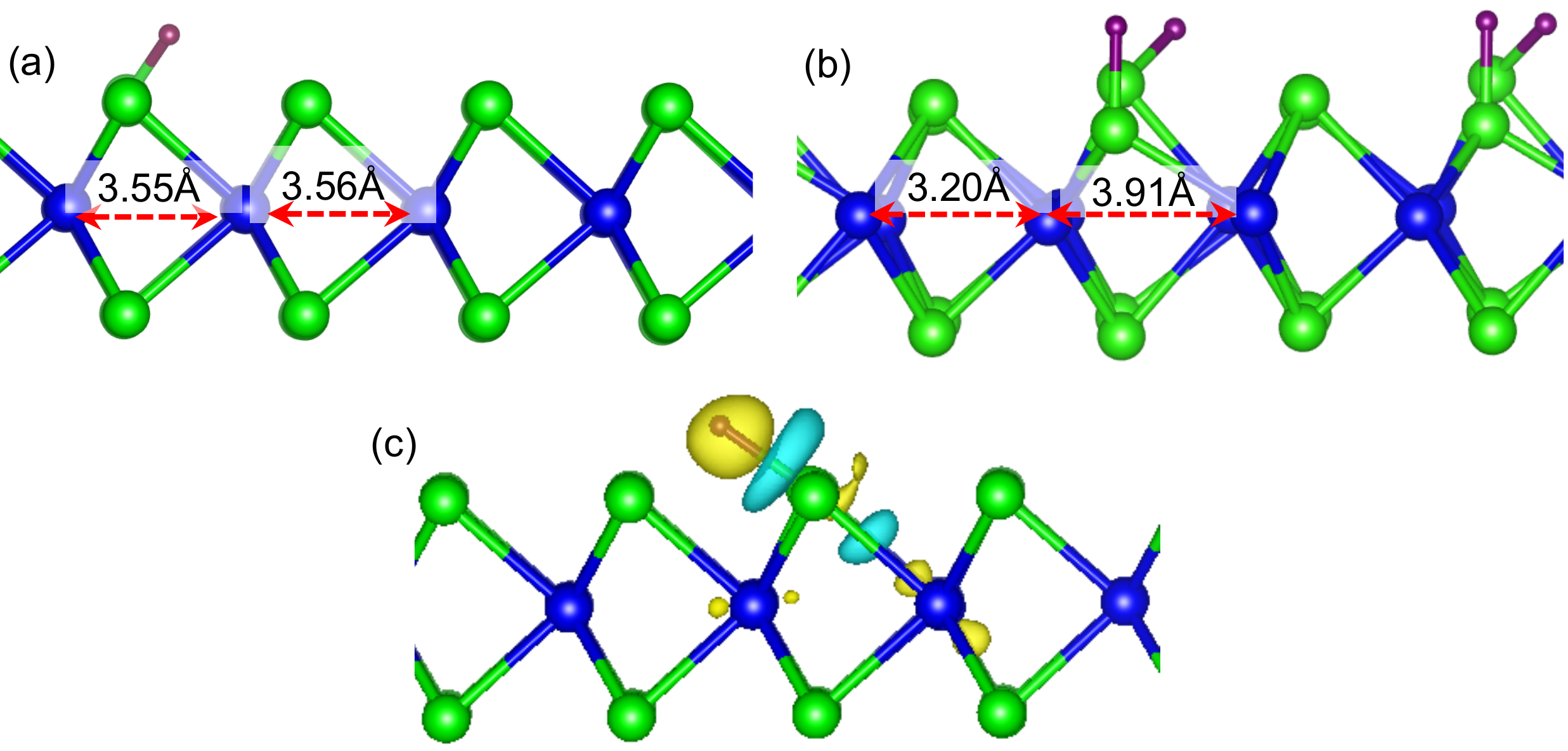}
	\caption{Hydrogenation-induced structural and electronic changes in $H$-phase MoTe$_{2}$ monolayer when the hydrogen coverage is (a) $3.12\,\%$ and (b) $50\,\%$. (c) Charge density difference for the  $3.12\,\%$ coverage, showing how hydrogenation depletes charge from Te-Mo bond (blue color), weakening the bond, which in turn leads to structural distortion.}
	\label{fig:structural-distortion}	
\end{figure}

The aforementioned shift in relative stability of the $H$- and $T^\prime$-phases can be related to the structural changes in the $H$-phase monolayer upon hydrogen-loading that herald the phase change within the structure. Fig.~\ref{fig:structural-distortion} shows the effects of hydrogen coverage on the structural arrangement of atoms within the MoTe$_{2}$ monolayers. For smaller hydrogen coverages, the distortion induced is small [Fig.~\ref{fig:structural-distortion}(a)]. However, the distortion within the $H$-phase MoTe$_{2}$ monolayer increases as a function of hydrogen coverage. For the higher coverages, a larger structural distortion results in the dimerization of Mo atoms as illustrated in Fig.~\ref{fig:structural-distortion}(b), with alternating long and short Mo-Mo distances resembling the $T^\prime$ phase. 
\newline 
The structural modifications within functionalized MoTe$_{2}$ monolayers that lead to the phase transition have associated electronic structural changes. Figure \ref{fig:structural-distortion}(c) shows the charge density differences for the hydrogen coverage of $3.12\,\%$ calculated using the following formula:
\begin{equation}
\Delta\rho= \rho\textrm{(tot)}-\rho\textrm{(MoTe$_{2}$)}-\rho\textrm{(H)}
\end{equation}
Here, $\rho$(tot) is the total charge density of the system, $\rho$(MoTe$_{2}$) is the charge density of the MoTe$_{2}$ monolayer and $\rho$(H) is the charge density of hydrogen atom. As shown in Fig~\ref{fig:structural-distortion}(c), in the $H$-phase at low concentration, most of the charge is accumulated (denoted by yellow color) on the hydrogen atom, and some is on the Mo-atom, whereas the charge is depleted around Mo-Te bond (blue color) even at such low concentration. This charge depletion from the Mo-Te bond weakens the structure, resulting in the observed structural distortion, which leads to Mo-Mo dimerization. Hence, we find that the changes in electronic structure upon hydrogenation, result in the observed local structural distortion. These structural distortions are suspected of further driving the redistribution of charges, resulting in phase change.

\begin{figure}
	\centering
	\includegraphics[width=0.35\textwidth]{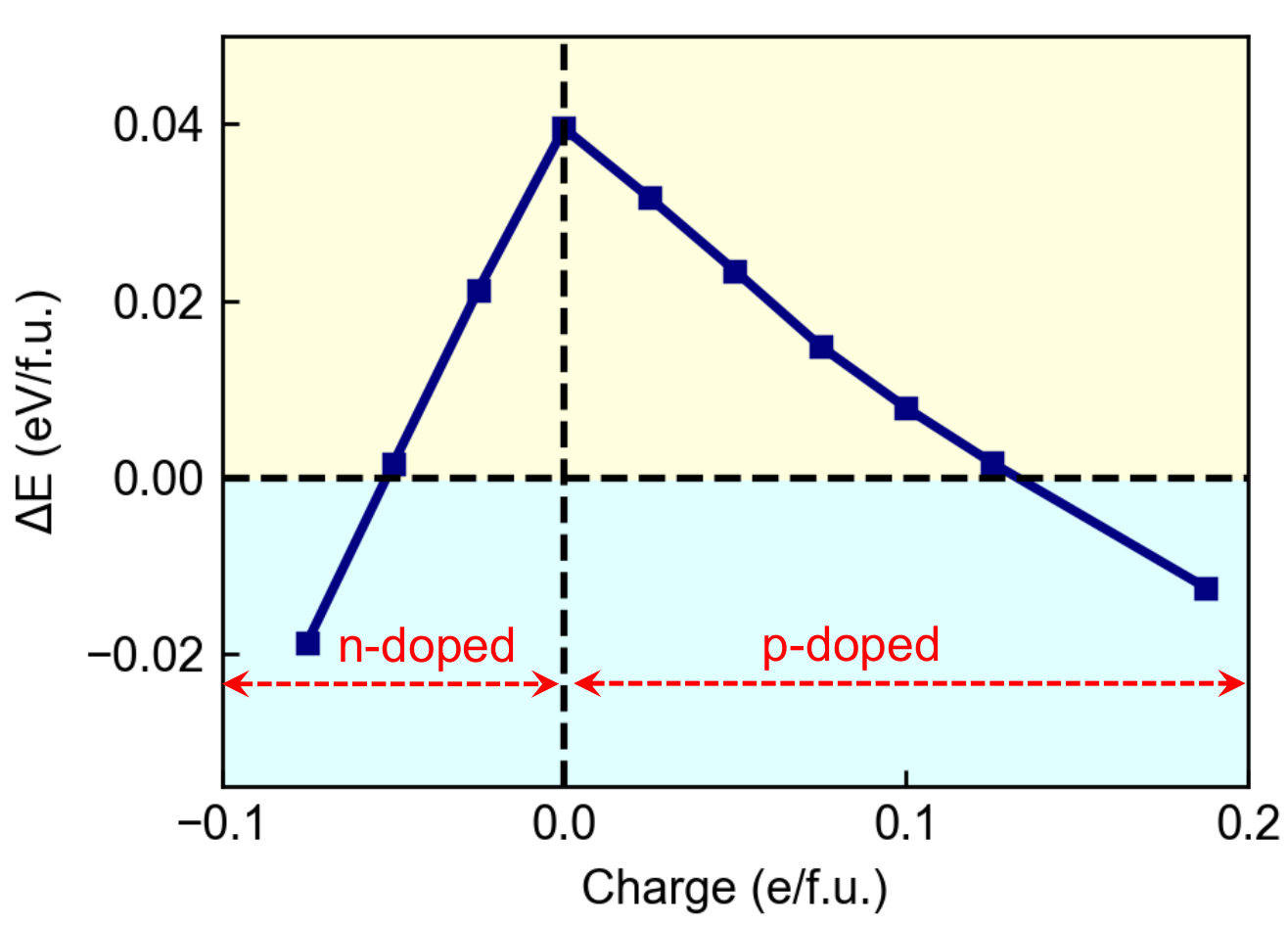}
	\caption{Energy difference between for $T^\prime$- and $H$-phases of MoTe$_{2}$ monolayers ($\bigtriangleup$E = E$^{T^\prime}$-E\textsuperscript{H}) as a function of charge-doping.}
	\label{fig:figure6}
\end{figure}

The proposed mechanism of phase change in functionalized MoTe$_{2}$ monolayers is also supported by the Bader charge analysis~\cite{Henkelman2006}, which indicates that hydrogen indeed hole-dopes the MoTe$_{2}$ layers. Our calculations show that the phase transition from the $H$ to $T^\prime$ phase occurs when at least 0.011\,e/f.u. charge is transferred from the MoTe$_{2}$ monolayer to the hydrogen atom. To determine if mere charging of the MoTe$_{2}$ monolayer is itself responsible for the phase change, we calculated the effect of simply adding charges to the pristine monolayers in the two phases. The calculated energy difference between the $T^{\prime}$ phase and the $H$-phase is plotted in Fig.~\ref{fig:figure6} as a function of added charge. It is clear that both electron- and hole-doping trigger the phase transformation. For the electron-doped system, the phase change occurs when the added charge is a $\sim0.050$\,e/f.u.. On the hand, employing hole-doping to trigger phase change (from $H$ to $T^\prime$) requires a larger doping level ($\sim0.132$\,e/f.u.). Thus, the amount of charge required to change the phase of a pristine MoTe$_{2}$ monolayer (by, say, gating it) is much larger than the charge-transfer calculated for the hydrogenated system. This suggests that the origin of phase transition in a hydrogenated MoTe$_{2}$ monolayer is not merely electrostatic doping. This is in contrast to alkali metal adsorption, where the adsorbate donates the charges to TMD layers, resulting in a phase change. The origin of the phase transition upon adsorption of alkali metal was found to be purely electrostatic doping~\cite{NasrEsfahani2015}. \newline

\begin{figure}
	\centering
	\hspace*{-0.5 cm}
	\includegraphics[width=0.35\textwidth]{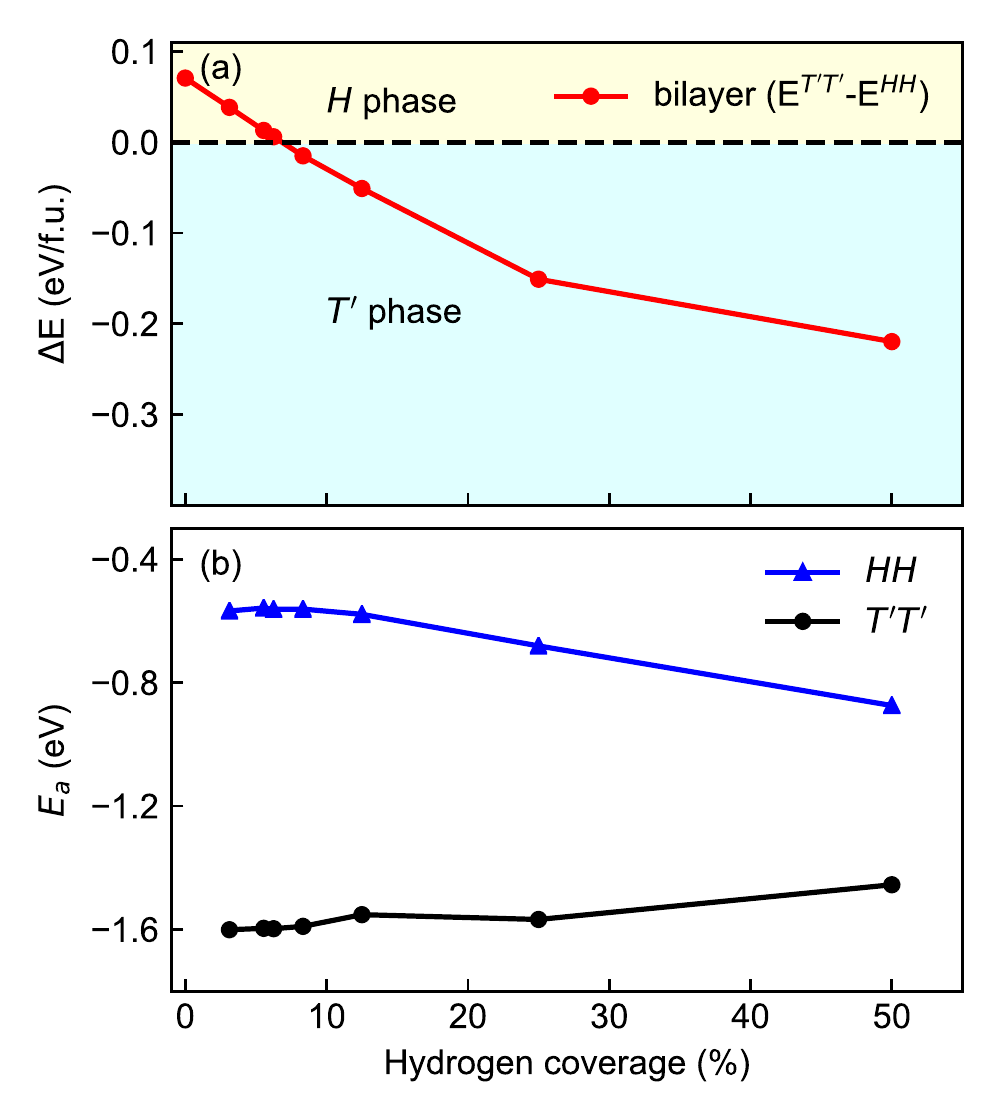}
	\caption{(a) The energy difference between different phases of MoTe$_{2}$ bilayers as a function of hydrogen concentration; (b) adsorption energy as a function of hydrogen concentration for $HH$ and $T^{\prime}T^{\prime}$ phases of MoTe$_{2}$ bilayers.}
	\label{fig:abs_bilayer}
\end{figure}

\noindent \textbf{TMD Bilayer:}  In order to investigate how the phase-change properties of MoTe$_2$ bilayers differ from those of the monolayers, we again considered a number of adatom coverages.  We calculated the total energy difference per formula unit ($\Delta$E) between bilayers in the $HH$ and $T^{\prime}T^{\prime}$ phases as a function of hydrogen coverage to find the relative stability of MoTe$_{2}$ bilayers in different phases. The percentage coverage at which the phase transition occurs is different for the bilayers as compared to the monolayers. This can be seen in Fig. \ref{fig:abs_bilayer}(a), which shows that we need to adsorb almost twice as much hydrogen on the upper layer of the composite structure for a phase transition from the $HH$-phase to the $T^{\prime}T^{\prime}$-phase. Our calculations also showed that for all coverages, the  $T^\prime$$T^\prime$ bilayer was more stable as compared to the $T^{\prime}H$ bilayer (not shown in the figure).
 
\begin{figure}
	\centering
	\hspace*{-0.5 cm}
	\includegraphics[width=0.4\textwidth]{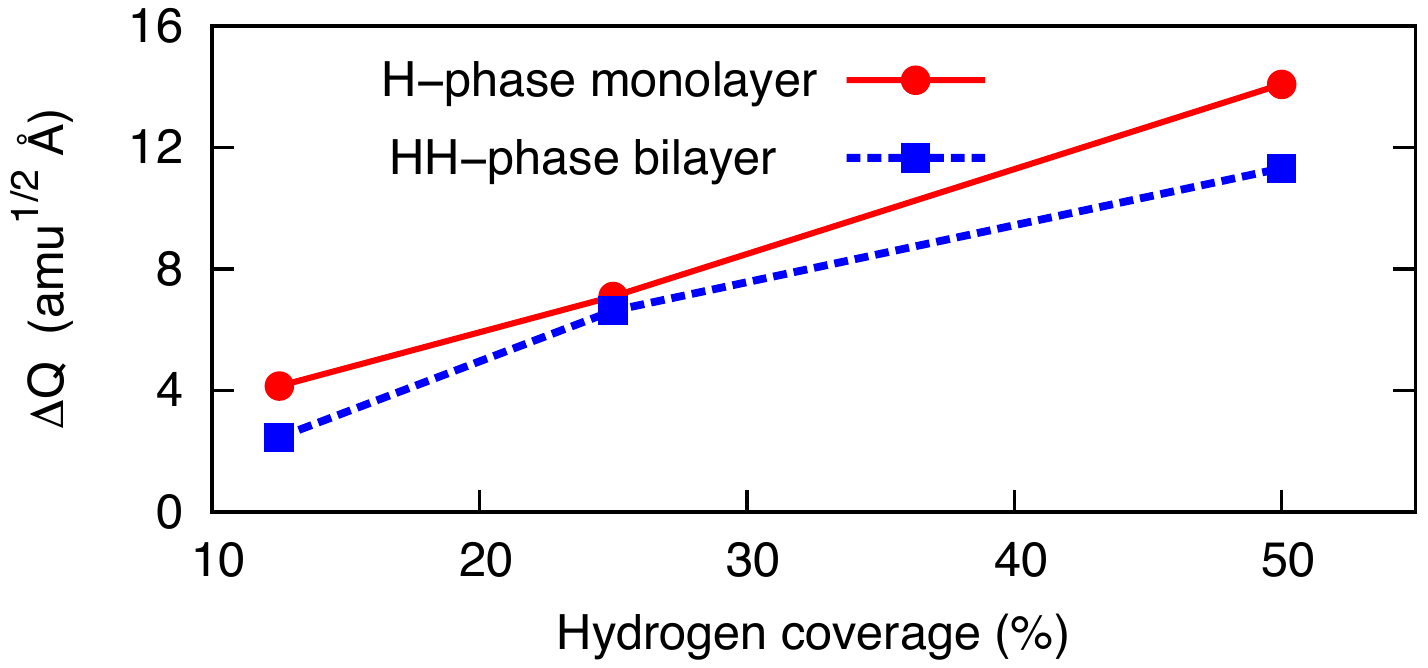}
	\caption{The change in geometry between pristine and hydrogenated layers of MoTe$_2$ as quantified by the change in generalized configuration coordinate, $\Delta \mathrm{Q}$, and plotted as a function of hydrogen coverage.}
	\label{fig:deltaQ}
\end{figure}

As seen in Fig.~\ref{fig:abs_bilayer}(b), bilayers show a very similar trend in the adsorption energies of hydrogen [calculated using equation~(\ref{equation:eq1})] compared to that seen in monolayers. This is not surprising as we are adsorbing hydrogen on the upper layer, and our choice of MoTe$_{2}$ as a substrate/lower layer within the composite structure ensures that the interfacial interactions are mostly van der Waals in nature. Surprisingly, even for layers bound by van der Waals interactions, interfacial interactions play a role when the top layer undergoes structural distortions upon hydrogenation that herald its phase change. The effect of substrate friction can be quantified by calculating the change in the generalized configuration coordinate, $\Delta \mathrm{Q}$, given by the formula:
\begin{equation}
 \Delta \mathrm{Q}^{2}=\sum_{i\alpha}\,m_{\alpha}(R_{i\alpha}^{EGH} - R_{i\alpha}^{EGP})^2
 \end{equation}
where, $m_{\alpha}$ is the atomic mass of the $\alpha^{th}$ atom, $R_{i\alpha}^{EGH}$ refers to the equilibrium geometry coordinates of the $\alpha^{th}$ atom  in the $i^{th}$-direction ($i={x,y,z}$) within the layer upon hydrogenation and $R_{i\alpha}^{EGP}$ refers to the corresponding equilibrium geometry coordinates of the $\alpha^{th}$ atom in the pristine layer. 
Fig.~\ref{fig:deltaQ} is a plot of $\Delta \mathrm{Q}$ for three representative hydrogen coverages, which were chosen because the supercell size for all three of them was the same. $\Delta \mathrm{Q}$ for each hydrogen coverage is larger in the freestanding monolayers of $H$-phase MoTe$_2$ as compared to those in bilayers, implying smaller changes in the geometry in the layer within a composite. Hence, a bilayer requires more hydrogen loading to affect the phase change. 

\section{Kinetics of Phase Change}

\begin{figure}[h]
	\centering
	\hspace*{-0.5 cm}
	\includegraphics[width=0.49\textwidth]{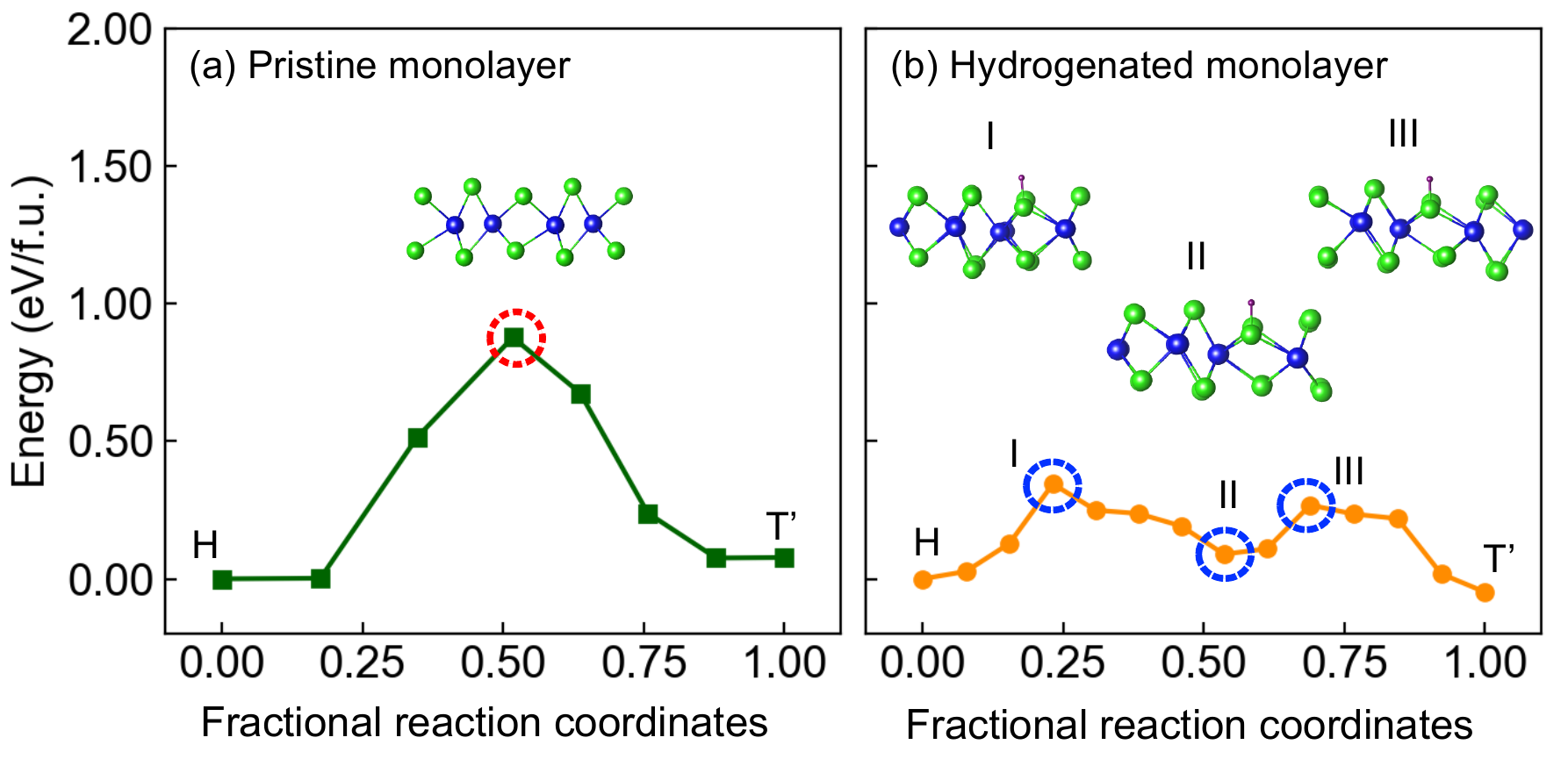}
	\caption{Transition energy barrier for the $H$ $\rightarrow$ $T^\prime$-phase transformation calculated using the nudged elastic band (NEB) method for (a) pristine MoTe$_{2}$ monolayer and (b) hydrogenated MoTe$_{2}$. The geometries adopted by the states corresponding to saddle points and stable intermediate states (encircled points) are also shown.}
	\label{fig:neb-monolayer}
\end{figure}

Our results presented thus far suggest that beyond a certain percentage of hydrogen coverage ($\sim4\%$ for monolayers and $\sim8\%$ for bilayers), the $T^\prime$-phase is thermodynamically more stable as compared to the $H$-phase. In order to determine the ease of transition between the two phases, we calculated the transition energy barriers using the nudged elastic band (NEB) method. The energy profile for a freestanding pristine MoTe$_{2}$ monolayer calculated using the NEB method, plotted in Fig.~\ref{fig:neb-monolayer}(a), shows a barrier between $H$ and $T^\prime$ phases of 0.795\,eV/f.u., consistent with previously calculated values~\cite{Zhang2016,Krishnamoorthy2018}. In order to study how the hydrogenation of MoTe$_2$ affects the kinetics of phase change from $H$ $\rightarrow$ $T^\prime$, we chose a hydrogen coverage at which $T^\prime$ is the lower energy structure ($12.5\%$). Fig.~\ref{fig:neb-monolayer}(b) shows the energy profile of the structural phase transition for a hydrogenated monolayer at this coverage, consisting of two saddle points and a stable intermediate state.  The saddle points and the intermediate low-energy state consist of different admixtures of the $H$- and $T^\prime$-phases, which are also shown in Fig.~\ref{fig:neb-monolayer}(b). The primary energy barrier has a value of to 0.340\,eV/f.u. [Fig.~\ref{fig:neb-monolayer}(b)]. Hence, the barrier energy associated with the hydrogenation-induced phase change is much smaller than the barrier calculated for the pristine monolayer. This is also much smaller than the barrier energies calculated for charge-doped (electron or hole) MoTe$_{2}$ monolayers~\cite{Zhang2016}. Our NEB results further confirm previously presented conclusion that the hydrogenation of MoTe$_{2}$ results in more than electrostatic doping of the layer.

\begin{figure*}[ht]
	\includegraphics[width=0.9\textwidth]{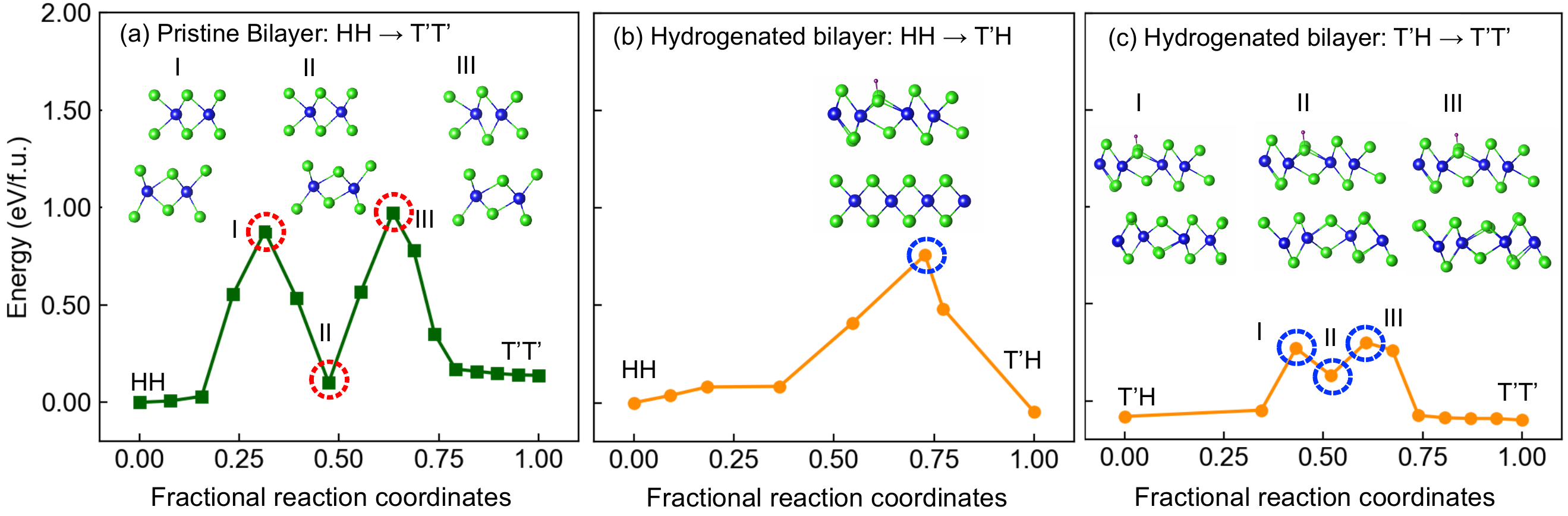}
	\caption{Energy barriers and pathways for transition from $HH$ $\rightarrow$ $T^{\prime}T^{\prime}$-phase bilayers using the nudged elastic band (NEB) method: (a) without hydrogenation, (b) with hydrogen adsorption between the initial state and the intermediate state ($HH$ $\rightarrow$ $T^{\prime}H$), and (c) with hydrogen adsorption between the intermediate state and the final state ($T^{\prime}H$ $\rightarrow$ $T^{\prime}T^{\prime}$).}
	\label{fig:neb-bilayer}
\end{figure*}

For the pristine MoTe$_{2}$ bilayer, the computed barrier between $HH$ $\rightarrow$ $T^{\prime}T^{\prime}$ is 0.975\,eV/f.u. [Fig.~\ref{fig:neb-bilayer}(a)]. The transition happens through the intermediate hybrid $T^{\prime}H$ phase as shown in Fig.~\ref{fig:neb-bilayer}(a). It is reasonable to assume that the hybrid $T^{\prime} H$ phase is an intermediate-state for the hydrogenated bilayers as well. Hence, we calculated the barrier in two parts: (i) bilayer $HH$ $\rightarrow$ $T^{\prime}H$ phase, and (ii)  $T^{\prime}H$ $\rightarrow$ $T^{\prime}T^{\prime}$. Fig.~\ref{fig:neb-bilayer}(b) shows the energy profile for the transition from the initial to the intermediate state, while Fig.~\ref{fig:neb-bilayer}(c) shows the pathway from the intermediate state to the final state. We find that for the hydrogenated MoTe$_{2}$ bilayer, the primary transition barrier is 0.760\,eV/f.u., which is only sightly smaller than that computed for pristine bilayers. This is in direct contrast with the results obtained for the monolayer where the energy barrier is significantly reduced upon hydrogenation. \newline

According to transition-state theory, the transition rate is estimated by $1/\tau = \nu e^{-E_b/kT}$, where $\tau$ is the transition time, $\nu$ is the attempt frequency, $E_b$ is the transition barrier energy (per formula unit), $k$ is the Boltzmann constant, and $T$ is the temperature (298 K). The attempt frequency is set at 10\,THz, which is the same order of magnitude as the frequencies for the optical phonons in $T^\prime$ monolayers~\cite{Qian2014}. The calculated phase-transition time for the pristine monolayer is $\sim 2.79$\,s, which reduced to 67.8\,ns with hydrogenation of the monolayer. For the bilayer, the transition barrier of 0.975\,eV/f.u. corresponds to a transition time of $\sim2.93\times 10^3$\,s, which reduced to about 0.67\,s upon hydrogenation. The latter is about $10^7$-times the timescale that one would predict, if considering a freestanding monolayer. Therefore, we find that that even the presence of a weakly interacting substrate, such as the one considered here (another MoTe$_2$ layer), affects every aspect of the the phase-change properties, including the timescales involved in the phase transition. 


\section{Conclusions}
In this theoretical work, we showed that the phase-change properties of TMD layers are modified by their immediate environment, by considering free-standing MoTe$_{2}$ and MoTe$_{2}$ placed on a substrate (another MoTe$_{2}$ layer). We used hydrogenation as a means to affect phase transition in MoTe$_{2}$.  In this proof-of-principle study, we show that for monolayers of MoTe$_{2}$, hydrogenation not only stabilizes the $T^{\prime}$ phase, but it also significantly reduces the transition energy barrier, thus increasing the $H \rightarrow T^{\prime}$ transition rate. In the case of bilayers, one needs a higher hydrogen coverage for a phase transition to occur. In addition, the kinetic energy barrier for the phase transition is considerably higher in hydrogenated bilayers than in the hydrogenated monolayers, with the associated timescale for a transition in the bilayer being about $10^7$-times longer than that in a monolayer at room temperature. The differences in properties of a monolayer in vacuum as compared to its properties when part of a composite will have to be taken into account when designing experiments and/or devices. Through inclusion of more realistic conditions, this work represents a critical step towards realizing full potential of layered materials in different applications.

\begin{acknowledgments}

This work is supported by NSF Grant number DMR-1752840. PD and PM acknowledge the computational support provided by the Extreme Science and Engineering Discovery Environment (XSEDE) under Project PHY180014, which is supported by National Science Foundation grant number ACI-1548562. For three-dimensional visualization of crystals and volumetric data, use of VESTA 3 software is acknowledged. 

\end{acknowledgments}

\bibliography{bibliography}
\bibliographystyle{apsrev}

\end{document}